\documentclass[prl,aps,twocolumn,showpacs]{revtex4-1} %floatfix
\usepackage{amsmath,amssymb,amsfonts,amsthm}
\usepackage{mathrsfs}
\usepackage{extarrows}
\usepackage{graphicx}
\usepackage{subfigure}
\usepackage{bm}
\usepackage{float}
\usepackage{color}
\usepackage{sidecap}
\allowdisplaybreaks
\usepackage{soul}
\setstcolor{red}
\usepackage[normalem]{ulem}
\usepackage{hyperref}
\hypersetup{colorlinks=true,breaklinks,urlcolor=blue,linkcolor=blue,citecolor=blue}

\begin{document}

\title{Topological Superconductors in One-Dimensional Mosaic Lattices}
\author{Qi-Bo Zeng}
\email{zengqibo@cnu.edu.cn}
\affiliation{Department of Physics, Capital Normal University, Beijing, 100048, China}
\author{Rong L\"u}
\author{Li You}
\affiliation{State Key Laboratory of Low Dimensional Quantum Physics, Department of Physics, Tsinghua University, Beijing 100084, China}
\affiliation{Frontier Science Center for Quantum Information, Beijing, China}
\begin{abstract}
We study topological superconductor in one-dimensional (1D) mosaic lattice whose on-site potentials are modulated for equally spaced sites. When the system is topologically nontrivial, Majorana zero modes appear at the two ends of the 1D lattice. By calculating energy spectra and topological invariant of the system, we find the interval of the mosaic modulation of the on-site potential, whether it is periodic, quasiperiodic, or randomly distributed, can influence the topological properties significantly. For even interval of the mosaic potential, the system will always exist in the topological superconducting phase for any finite on-site potentials. When the interval is odd, the system undergoes a topological phase transition and enters into the trivial phase as the on-site potentials become stronger than a critical value, except for some special cases in the commensurate lattices. These conclusions are proven and the phase boundaries determined analytically by exploiting the method of transfer matrix. They reveal that robust Majorana zero modes can arise in 1D mosaic lattice independent of the strength of the spatially modulated potentials.
\end{abstract}
\maketitle
\date{today}

\emph{Introduction}.---During the past decade, topological superconductor (TSC) becomes a topic of extensive study~\cite{Alicea2012PRP,Beenakker2013ARCMP,Elliott2015RMP,Ando2015ARCMP,Sato2017RPP}. Under appropriate conditions, a TSC will become topologically nontrivial and hosts Majorana fermions, or Majorana zero modes (MZMs), which play an important role in implementing topological quantum computation~\cite{Sarma2015npj,Obrien2018PRL,Lian2018PNAS,Litinski2018PRB}. Many theoretical schemes for realizing MZMs are proposed in various platforms, ranging from one-dimensional $p$-wave superconductor~\cite{Lutchyn2010PRL,Oreg2010PRL}, to two-dimensional $p+ip$ superconductors~\cite{Stone2004PRB,Fendley2007PRB,Raghu2010PRL,Chung2011PRB}, superfluid He-3~\cite{Kopnin1991PRB,Qi2009PRL,Chung2009PRL}, ultracold atoms~\cite{Liu2012PRA,Qu2013NatCom,Chen2013PRL,Ruhman2015PRL}, and magnetic atom chain on superconducting substrates~\cite{Perge2013PRB,Hui2015SR,Dumitrescu2015PRB}. Among all these proposals, the search for MZMs in one-dimensional (1D) semiconductor-superconductor heterostructures has attracted the most attention both theoretically~\cite{Ioselevich2011PRL,Zazunov2011PRB,Wu2012PRB,Jose2012PRL,Ueda2014PRB,Zeng2016FP}, and experimentally~\cite{Mourik2012Science,Das2012NatPhys,Finck2013PRL,Perge2014Science}.

The prototype lattice model for describing 1D TSC is put forward by Kitaev~\cite{Kitaev2001}, where uniform on-site potentials are imposed on every lattice site. However, disorders ubiquitously exists in real experiments and materials. Random disorders will localize all the eigenstates in 1D lattices, commonly refereed to as Anderson localization (AL) phenomenon~\cite{Anderson1958PR}. In a TSC, disorders are detrimental to the existence of topological phase~\cite{Motrunich2001PRB,Potter2010PRL,Shivamoggi2010PRB,Akhmerov2011PRL,Brouwer2011PRL,Lobos2012PRL,Sau2012NC,Adagideli2014PRB}. Besides, correlated disorders such as a quasiperiodic potential can also destroy TSC due to localization phase transition~\cite{Cai2013PRL,DeGottardi2013PRL,Zeng2016PRB}. To facilitate robust applications in topological quantum computation, it would be essential to construct TSC that is insensitive to spatially varying potentials. Since MZMs lie in the band gap around zero energy, one may expect their appearance if the band gap is protected from AL. One way to separate localized or extended states is by introducing mobility edges. Recently, a new type of 1D lattice, which is called mosaic lattice, is investigated~\cite{Wang2020arxiv}, with quasiperiodic potentials are only added onto equally spaced sites, giving rise to multiple mobility edges. Moreover, in some cases, the mobility edges will stay at zero energy unaffected by AL unless the quasiperiodic potential is infinitely strong. It is imperative to ask whether similar mechanisms can be used to protect MZMs from strong disorders.

In this paper, we generalize the one-dimensional topological superconductor model by introducing mosaic modulations to on-site potentials, which are added to equally spaced sites. These potentials can be either periodic, quasiperiodic, or randomly distributed. The topological properties of such 1D TSC are shown to be influenced significantly by the interplay between mosaic modulation and superconducting pairing. When the interval of the mosaic modulation is even, we find the band gap around zero energy in the energy spectra persists, no matter how strong the on-site potentials are. Thus, the system is always topologically nontrivial and MZMs can be observed at any finite values of the on-site potentials. However, if the interval of the mosaic potentials becomes odd, the system will enter into the trivial phase as the strength of the on-site potential increases, except for a few special cases in the commensurate lattices. We prove these exotic results analytically by making use of the transfer matrix method, which is also employed to calculate the $Z_2$ topological invariant and determine the phase boundaries of topological phase transitions. Our work reveals that in the 1D mosaic lattices, one can construct robust TSC and obtain MZMs that survive arbitrary spatially varying potentials, and will support promising applications in realizing topological quantum computation.

\emph{Model Hamiltonian}.---We consider the one-dimensional mosaic lattice model with $p$-wave superconducting pairing which is described as
\begin{equation}\label{H}
H = \sum_{j=1}^{L-1} [-t c_{j+1}^\dagger c_j + \Delta c_{j+1}^\dagger c_j^\dagger + H.c.] + \sum_{j=1}^{L} V_j c_j^\dagger c_j,
\end{equation}
where $c_j^\dagger$ ($c_j$) is the creation (annihilation) operator of spinless fermion at site $j$. $t$ is the hopping amplitude between the nearest neighboring lattice sites and we take $t=1$ as the energy unit throughout this paper. $\Delta$ denotes the amplitude of the $p$-wave superconducting pairing in the system. The on-site potential $V_j$ is represented by the following function
\begin{equation}\label{V}
V_j = \left\lbrace 
\begin{aligned}
& V \cos[2 \pi (\alpha j + \phi)],\qquad j = s \kappa, \\
& 0,\qquad \text{otherwise}.
\end{aligned}
\right. 
\end{equation}
and $s=1,2,\cdots,N$. Thus, the on-site potential occurs with interval $\kappa$ and we can define a unit cell with the nearest $\kappa$ lattice sites. The system size becomes $L=\kappa N$ with $N$ the number of the unit cell. In addition, $\alpha$ determines the period of the potential and $\phi$ is a phase factor. If $\Delta=0$ and $\alpha$ is irrational, the model reduces to the mosaic model discussed in Ref. \cite{Wang2020arxiv}, where multiple mobility edges are reported. With the presence of $p$-wave superconducting (SC) pairing, there exists a band gap around zero energy. This work will focus on the interplay between SC pairing and mosaic modulation.

With the Bogoliubov-de Gennes (BdG) transformation $\eta_n^\dagger = \sum_{j=1}^{L}[ u_{n,j} c_{n,j}^\dagger + v_{n,j} c_{n,j} ]$, where $n$ refers to the $n$th eigenstate of the Hamiltonian, the eigenstate $|\Psi_n \rangle$ takes the form $|\Psi_n \rangle = \eta_n^\dagger |0 \rangle$, and satisfies the Schr\"odinger equation $H |\Psi_n \rangle = E_n |\Psi_n \rangle$. Representing the eigenstate as a 2L-dimensional column vector $|\Psi_n \rangle=[u_{n,1}, v_{n,1}, u_{n,2}, v_{n,2}, \cdots, u_{n,L}, v_{n,L}]^T$, the Hamiltonian can be expressed as a $2L \times 2L$ matrix (for details, see Ref.~\cite{SM}), which can be numerically diagonalized to obtain the eigenenergies and eigenstates.

We will investigate both the commensurate and incommensurate cases with $\alpha$ being rational or irrational numbers respectively. To distinguish the localized bulk states and Majorana zero modes from the extended bulk states, we define the inverse participation ratio (IPR) for each eigenstate $|\Psi \rangle$ as $\text{IPR} = \sum_j (u_{n,j}^4 + v_{n,j}^4)$, which is of a finite value $O(1)$ for a localized state. If the state is extended, the IPR will tend to be a miniscule value close to zero.  

\begin{figure}[t]
  \includegraphics[width=3.3in]{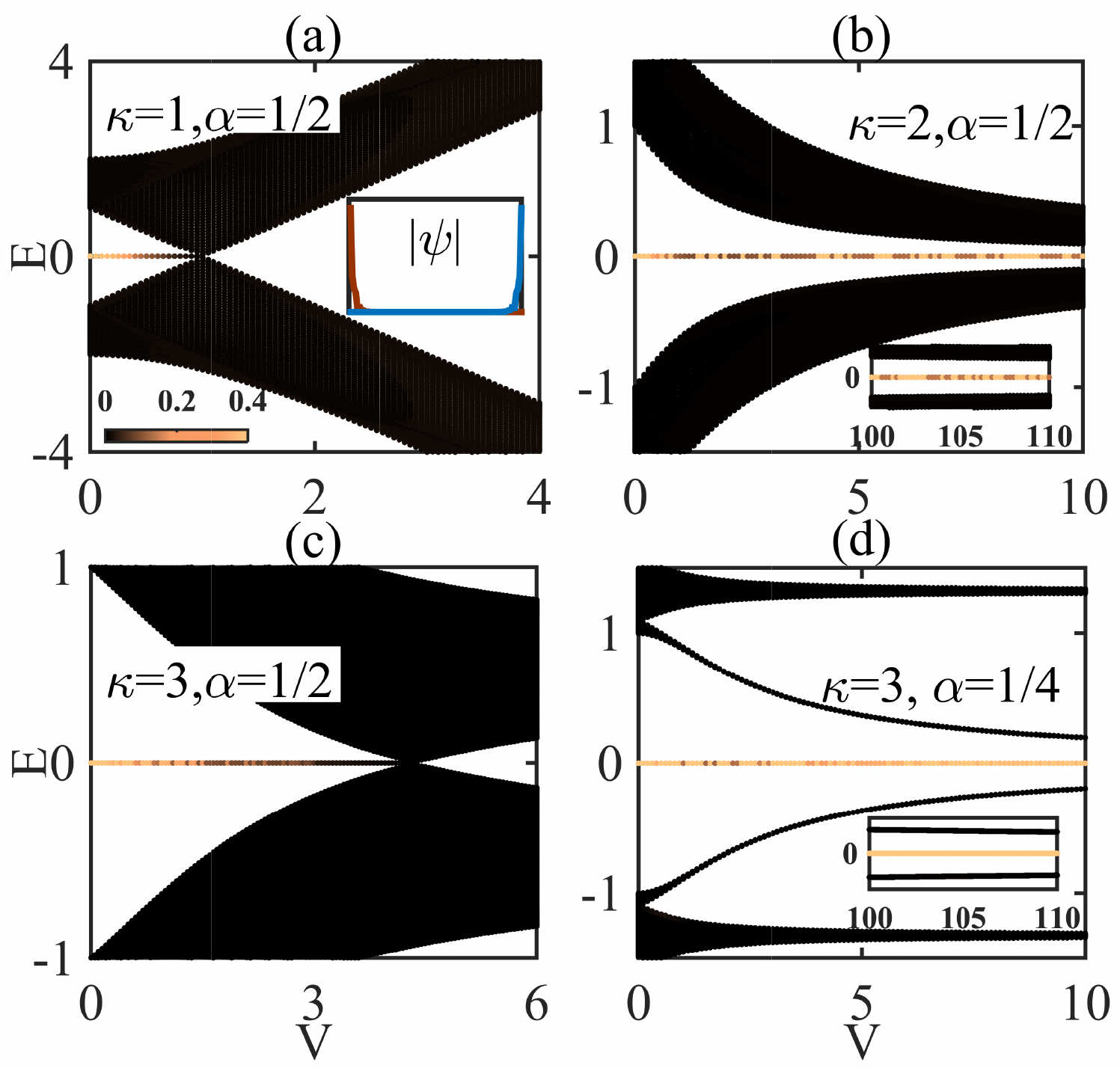}
  \caption{(Color online) Energy spectra for mosaic lattices as a function of on-site potential $V$ for $\alpha=1/2$ at (a) $\kappa=1$; (b) $\kappa=2$; (c) $\kappa=3$. While in (d) we set $\kappa=3$, $\alpha=1/4$. Color bar in (a) indicates the IPR for the corresponding eigenstate. The inset in (a) shows the Majorana zero modes localized at the two ends of the 1D lattice. Insets in (b) and (d) are for systems with stronger onsite potentials ($V \sim 100$), where the gap around zero energy still exists and thus the MZMs persist for even $\kappa$ at $\Delta=0.5$, $\phi=0$, $N=100$, and $L=\kappa N$.}
\label{fig1}
\end{figure} 

\emph{Commensurate lattice}.---We first check the properties for with commensurate modulation. In this case, $\alpha$ is a rational number that can be set to $\alpha=p/q$, with $p$ and $q$ co-prime integers. For the case with constant potential at each site, i.e. $\kappa=1$, $V_j=V$, it is well known that the model system is topologically nontrivial when $|V|<2t$. Majorana zero modes (MZMs) emerge in the energy gap in this regime. When the mosaic on-site potential is introduced. Figure \ref{fig1} shows the energy spectra of the Hamiltonian in Eq.~(\ref{H}) as a function of the on-site potential strength $V$, here for $\alpha=1/2$ and $\Delta=0.5$. When $\kappa=1$, we have a regular periodic lattice, the system is found to be topologically nontrivial when $|V|<1$. As can be seen in Fig. \ref{fig1}(a), there appear MZMs localized at the two ends of the lattice while all the bulk states are extended. When $V$ becomes stronger, the gap around zero energy becomes closed at $|V_c|=1$ and reopens when $|V|>1$ while the system enters the trivial phase with no edge modes present. Such a phenomenon is also observed when $\kappa=3$ and $\alpha=1/2$, except that the critical value of $V$ becomes different, as shown in Fig. \ref{fig1}(c). However, if we change $\alpha$ to $1/4$, the phase transition will not happen and the band gap around zero energy is finite even for much stronger on-site potential ($V \sim 100$), see Fig. \ref{fig1}(d). Meanwhile, MZMs reside in the gap. In fact, the system remains nontrivial unless the value of $V$ becomes infinite. Extensive numerical calculations indicate that, for the system with an odd $\kappa$, the above illustrated phenomenon always shows up when $q=2^k$ with $k \geq 2$.  
  
On the other hand, if $\kappa$ is even, the energy gap also persists no matter how strong the on-site potential is. For instance, in Fig. \ref{fig1}(b), we present the energy spectra for the mosaic lattice with $\kappa=2$. As $V$ increases, the gap around the zero energy becomes reduced, although it always remaining finite. For much stronger on-site potential, the band gap and thus the MZMs remain, as shown in the insets. The gap will only close in the limit when the on-site potential becomes infinitely strong. Otherwise, the system will always be topologically nontrivial and the MZMs are observed for any value of $V$. 

\begin{figure}[t]
  \includegraphics[width=3.4in]{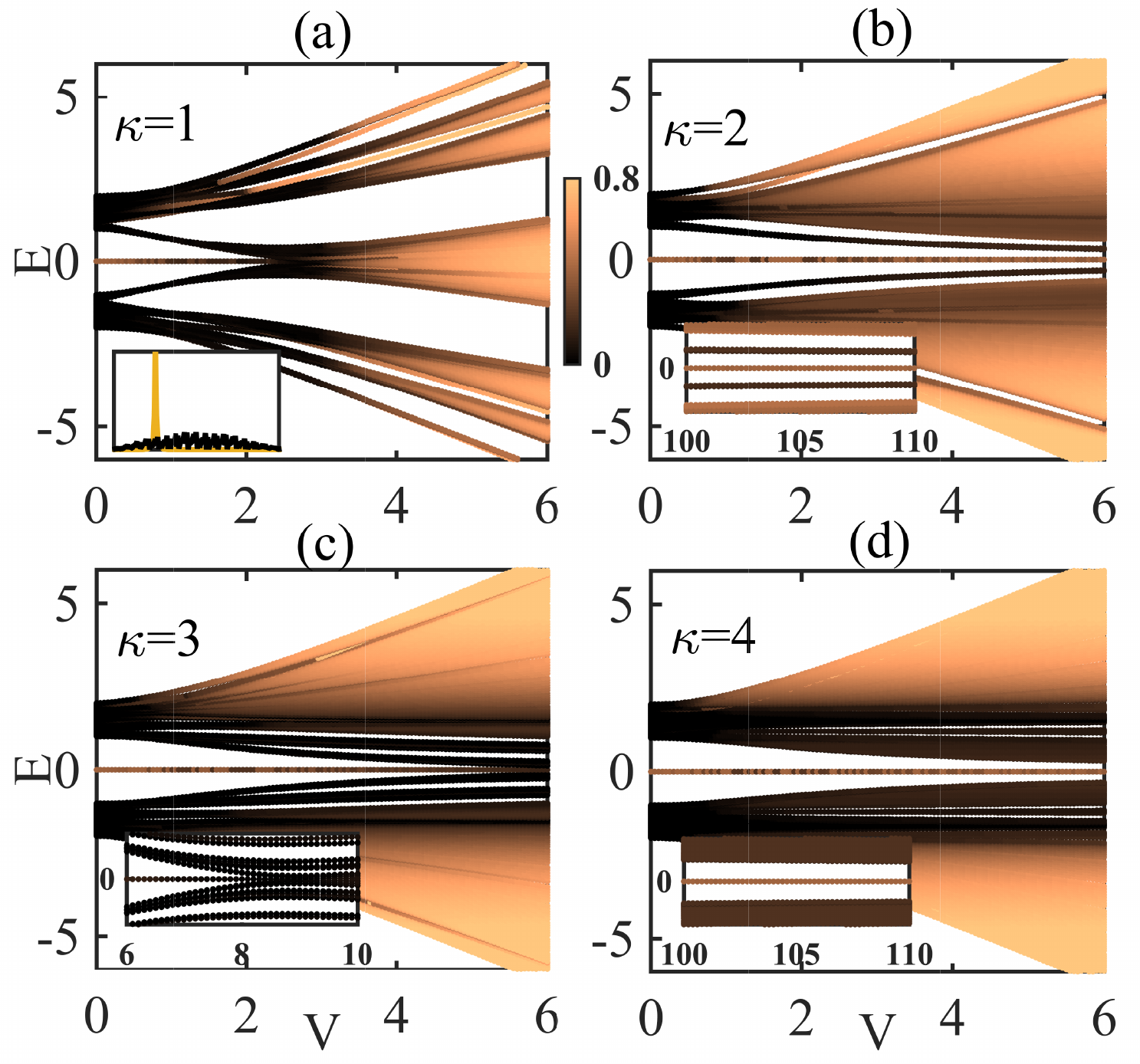}
  \caption{(Color online) Energy spectra for the 1D mosaic lattice with on-site modulation: (a) $\kappa=1$; (b) $\kappa=2$; (c) $\kappa=3$; and (d) $\kappa=4$, coded according to the color bar in (a) for the IPR for the corresponding  eigenstates. The inset in (a) illustrates the amplitudes of the localized and extended states. Insets in (b), (c) and (d) are the spectra for the system with stronger on-site potentials for $\Delta=0.5$, $\alpha=(\sqrt{5}-1)/2$, $\phi=0$, $N=144$, and $L=\kappa N$.}
\label{fig2}
\end{figure}

\emph{Incommensurate lattice}.---Next, we turn to the case of incommensurate lattice with $\alpha$ an irrational number. Without loss of generality, we set $\alpha=(\sqrt{5}-1)/2$. If $\kappa=1$, the mosaic lattice reduces to an usual quasiperiodic lattice, whose topological superconducting phase is destroyed by the AL phase transition at $V_c=2(t+\Delta)$~\cite{Cai2013PRL}. In Fig.~\ref{fig2}(a), we present the energy spectrum for $\kappa=1$. There exist no mobility edges, the bulk states are either extended or localized. When $V>2(t+\Delta)$, MZMs disappear and the system enters into the trivial phase. If $\kappa$ is different from 1, we expect mobility edges to appear since multiple mobility edges have already been found in the normal mosaic lattice with $\Delta=0$~\cite{Wang2020arxiv}. Figure~\ref{fig2}(c) shows the energy spectrum of the system for $\kappa=3$. From the IPR values of the eigenstates, we find that mobility edges can be observed as well, which separate the extended and localized states in the spectrum. Besides, as the on-site potential increases, more states will become localized. When $V$ is stronger than the critical value around $V=4.3$, the gap around zero energy will be closed and all MZMs disappear, i.e. the system becomes topologically trivial. However, the situation is totally different for even $\kappa$. Mobility edges can still be observed and almost all bulk states are localized when $V$ becomes strong enough, see Fig. \ref{fig2}(b) and (d). It is interesting to note that the energy gap around $E=0$ is always finite and MZMs can be observed in the gap. So even in the incommensurate case, mosaic lattices with even $\kappa$ only becomes trivial in the limit of infinitely strong on-site potential.

\begin{figure}[t]
  \includegraphics[width=3.4in]{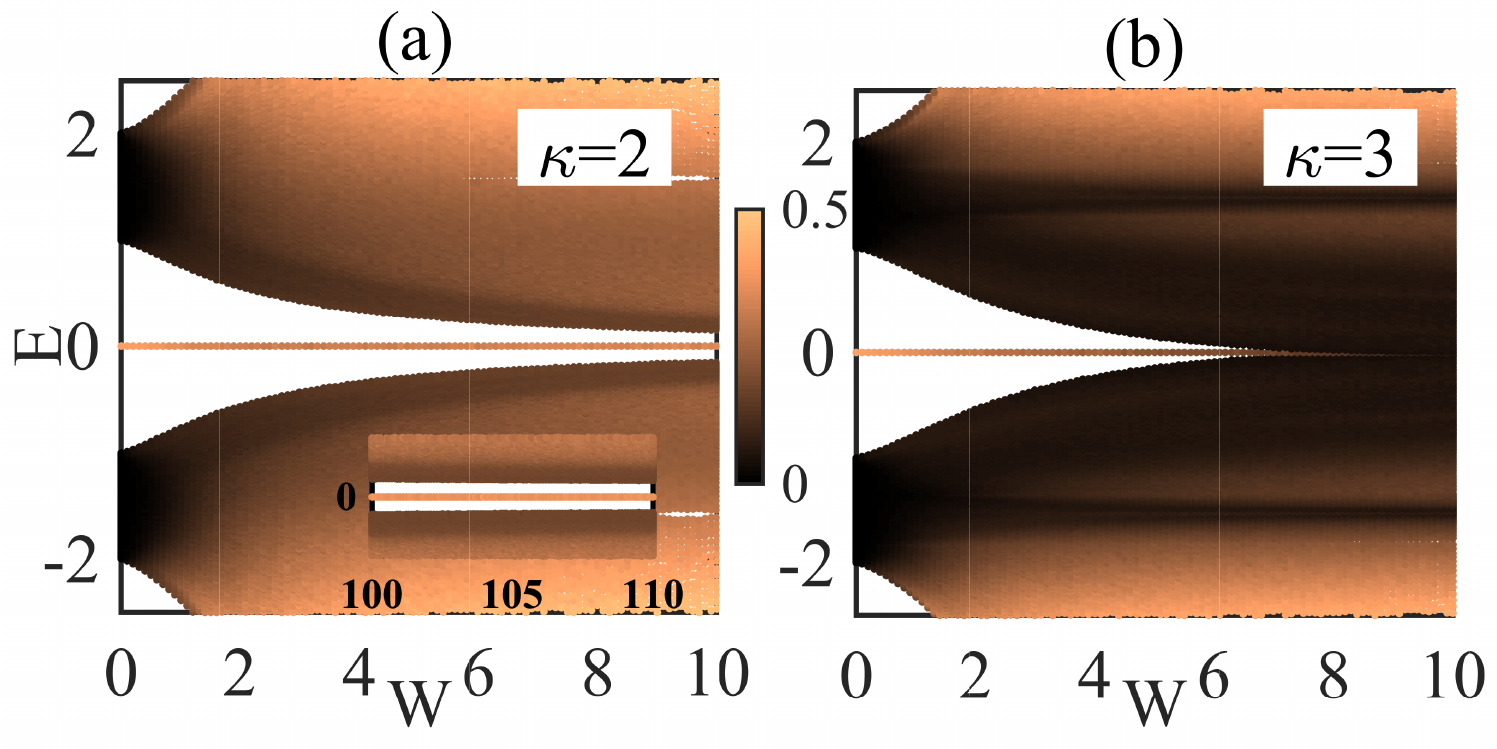}
  \caption{(Color online) Energy spectra for the 1D mosaic lattices with disordered on-site potentials: (a) $\kappa=2$; (b) $\kappa=3$ coded according to the color bar for the IPR of the eigenstate. The mosaic on-site potentials at the ($s\kappa$)th sites in the system are uniformly distributed in $[-W,W]$. The spectra are obtained by averaging over $100$ different samples for $\Delta=0.5$, $\phi=0$, $N=100$, and $L=\kappa N$. Inset shows result for stronger disorder.}
\label{fig3}
\end{figure}

\emph{Disordered lattice}.---Now we check the case of mosaic lattice with disordered on-site potential. In Fig.~\ref{fig3}, we present the energy spectra of disordered lattices. Here nonzero potentials at the ($s\kappa$)th sites are uniformly distributed in $[-W,W]$ and average over $100$ different samples are carried out for each $W$. Again, we are interested in finding out whether the topological phase transition happens or not as we tune the disordered potentials depends on the value of $\kappa$. Essentially the same is found as the periodic and quasiperiodic cases explored above, systems with even $\kappa$ will always be nontrivial while those with odd $\kappa$ will become trivial when the disorders become sufficiently strong.

From the discussions above, we find that the parameter $\kappa$, i.e., the spatial interval of the modulated potentials influences topological properties significantly. By choosing an even value of $\kappa$, we can obtain Majorana zero modes in the mosaic lattices at any finite on-site potential, whether it is commensurate, incommensurate, or randomly distributed.

\emph{Transfer matrix method and topological invariant}.---To further illuminate topological properties of the 1D mosaic lattice with $p$-wave pairing, we make use of the transfer matrix, as introduced in Refs.~\cite{DeGottardi2011NJP,DeGottardi2013PRL}. With $c_j=(a_j+ib_j)/2$, Dirac fermion can be expressed as the combination of two Majorana fermions, denoted by $a_j$ and $b_j$ at the $j$th site. These Majorana fermion operators satisfy the anticommutation rules $\{a_j,a_k\}=\{b_j,b_k\}=2\delta_{j,k}$ and $\{a_j,b_k\}=0$. The Hamiltonian in Eq.~(\ref{H}) is then rewritten in terms of majorana fermions as $H=\frac{i}{2}\sum_j [(-t+\Delta)a_j b_{j+1}+ (t+\Delta)b_j a_{j+1} - V_j a_j b_j]$. We further define the operators $Q_A=\sum_j \alpha_j a_j$ and $Q_B=\sum_j \beta_j b_j$ to represent the Majorana modes at the ends of the lattice. Then from the equation of motion, we find
\begin{equation}
\begin{pmatrix} \alpha_{j+1} \\ \alpha_j \end{pmatrix} = 
\begin{pmatrix} \frac{V_j}{t+\Delta} & \frac{\Delta-t}{\Delta+t} \\ 1 & 0 \end{pmatrix} 
\begin{pmatrix} \alpha_j \\ \alpha_{j-1} \end{pmatrix}
= A_j \begin{pmatrix} \alpha_j \\ \alpha_{j-1} \end{pmatrix},
\end{equation}
where $A_j$ is the transfer matrix for the $j$th site. As detailed in Ref.~\cite{DeGottardi2011NJP}, the existence of Majorana zero modes is determined by the eigenvalues of the transfer matrix of the whole lattice $\mathcal{A}=\prod_{j=1}^L A_j$. Furthermore, we can define a $Z_2$ topological invariant as $\nu=-\text{sgn}[f(1)f(-1)]$, where $f(z)=\text{det}(\mathcal{A}-Iz)$ is the characteristic polynomial of $\mathcal{A}$. The invariant $\nu=-1$ indicates a nontrivial phase while $\nu=1$ corresponds to a trivial phase.

With the method of transfer matrix and the topological invariant at hand, we explain the phenomena observed in the above based on the energy spectra analytically. We first check the mosaic lattice with even $\kappa=2m$ ($m=1,2,\cdots$). The transfer matrix for the whole lattice $\mathcal{A}$ is computed to be
\begin{equation}
\mathcal{A}=\begin{pmatrix}
\delta^{mN} & 0 \\ \frac{(\sum_{j=1}^L V_j) \delta^{mN-1}}{\Delta+t} & \delta^{mN}
\end{pmatrix},
\end{equation}
with $\delta=\frac{\Delta-t}{\Delta+t}$. In this case, the characteristic polynomial, as well as the topological invariant, will not depend on the on-site potential $V$. Through a simple calculation, we find $\nu$ is always fixed at $-1$, see Ref.~\cite{SM} for details. This implies that the system is always topologically nontrivial, no matter how strong the on-site potentials $V$ is, which is consistent with the result we obtain from the energy spectra shown in Fig.~\ref{fig1}. It is important to note this conclusion is independent of the specific form for on-site potential, which might be periodic, quasiperiodic, or randomly distributed. It applies to any mosaic lattice with even $\kappa$. Thus we analytically prove that topological superconducting phase persists in 1D mosaic lattices with any form or any strength of on-site potentials, as long as $\kappa$ is even.

However, if $\kappa$ is odd, the system becomes trivial as $V$ increases, except for the commensurate lattices with $q=2^k$ ($k \geq 2$) where nontrivial phase survives any finite on-site potential. We can use the transfer matrix method to prove this and determine the phase transition point analytically~\cite{SM}. For instance, at $\alpha=1/2$, $t=1$, and $\Delta=0.5$, we find the critical values for the phase transition are $|V_c|=1$ and $|V_c|=\frac{13}{3} \approx 4.3$ for  mosaic lattice with $\kappa=1$ and $3$, respectively. These results agree well with the gap closing points in the energy spectra. To determine the phase transition point in incommensurate mosaic lattice, we introduce Lyapunov exponent $\gamma(E)$, which is equal to the inverse of the localization length. In the model system with $p$-wave pairing it is related to $\gamma_0(E)$ for the normal model with no superconducting pairing~\cite{DeGottardi2013PRL}. For normal mosaic lattice, it is known that multiple mobility edges arise for $\kappa>1$~\cite{Wang2020arxiv}. Focusing on the Majorana zero modes, we only need to consider the behaviors of the zero-energy states. Thus we set $E=0$ in the formulation of Lyapunov exponent and we find
\begin{equation}
\gamma_0(0)= \left\{ \begin{aligned} 
& 0, \qquad \kappa \quad \text{even};\\
& \frac{\ln \frac{|V|}{2}}{\kappa}, \qquad \kappa \quad \text{odd}.
\end{aligned} \right.
\end{equation}
Hence, in incommensurate mosaic lattice with even $\kappa$, there exists no phase transition. The system always stays in the nontrivial phase where Majorana zero modes can be observed at the ends of the lattice. However, in lattice with odd $\kappa$, there exists a phase transition to AL. The phase boundary can be calculated to be 
\begin{equation}
|V_c|=2 \sqrt{\frac{(t+\Delta)^{\kappa+1}}{(t-\Delta)^{\kappa-1}}} \xlongequal{\kappa=2m+1} \frac{2(t+\Delta)^{m+1}}{(t-\Delta)^m},
\end{equation}
with integer $m$~\cite{SM}. Again, these agree well with the numerical results given in Fig.~\ref{fig2}.

\begin{figure}[t]
  \includegraphics[width=3.4in]{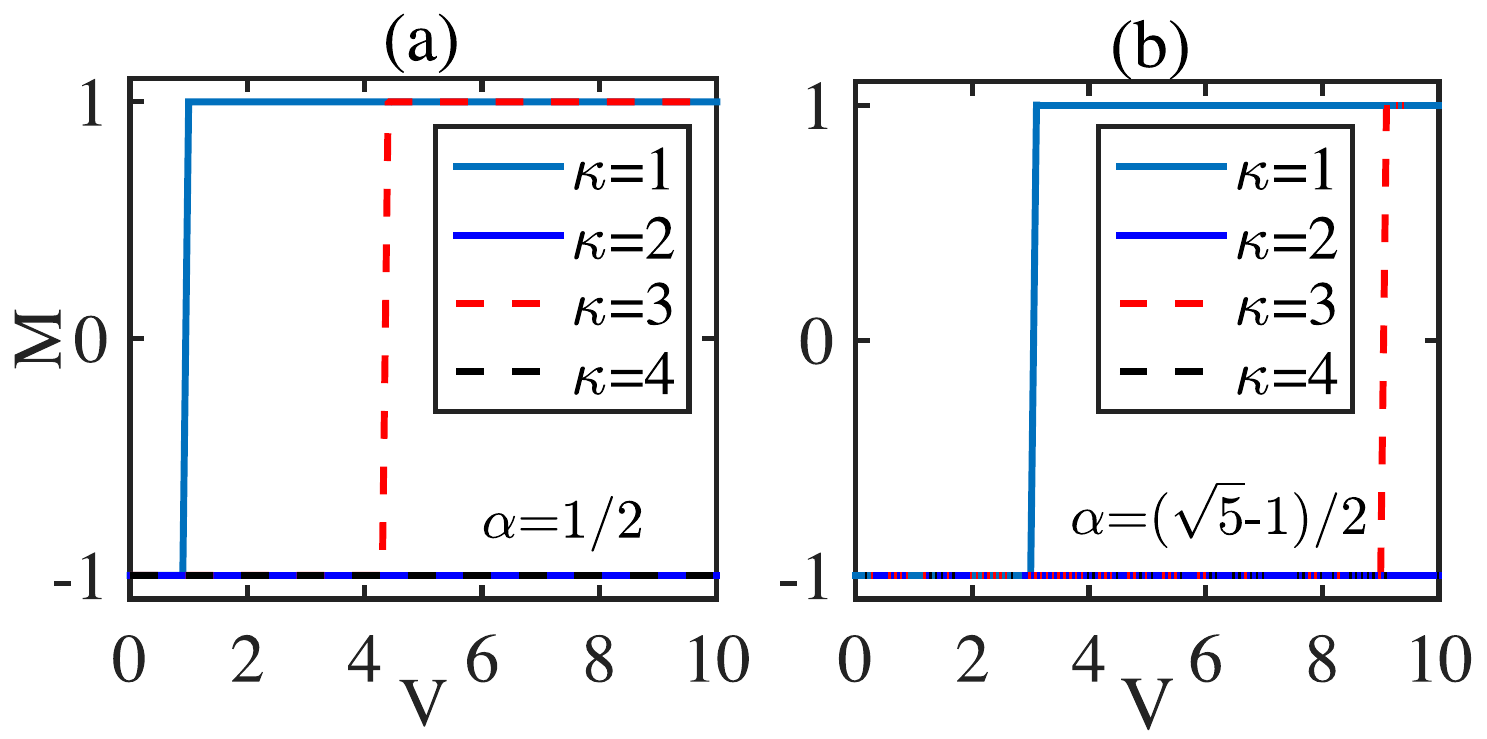}
  \caption{(Color online) $Z_2$ Topological invariant versus on-site potential $V$ for the 1D mosaic TSC system with (a) $\alpha=1/2$ and (b) $\alpha=(\sqrt{5}-1)/2$. $M=-1$ denotes topologically nontrivial phase while $M=1$ labels the trivial phase. Other parameters are $\Delta=0.5$, $\phi=0$, and $N=100$ in (a) and $144$ in (b). The lattice size is $L=\kappa N$.}
\label{fig4}
\end{figure}

An alternative approach to define a $Z_2$ topological invariant for the 1D topological superconductor is by using the Pfaffian of the system~\cite{Kitaev2001,Cai2013PRL}. The Hamiltonian in this case can be represented by skew-symmetric matrices in terms of Majorana operators, with the Pfaffian of the system is calculated using these matrices. The corresponding $Z_2$ topological invariant $M$ is determined by the sign of the Pfaffian. Figure~\ref{fig4} shows the numerical results of $M$ for both commensurate and incommensurate lattices. Again, $M=-1$ implicates topologically nontrivial phase. If the system enters into trivial phase, $M$ will jump from $-1$ to $+1$ at the critical on-site potential $V_c$. We find that for even $\kappa$, $M$ is always $-1$, which implies that the system is nontrivial irrespective of the value of $V$. If $\kappa$ is odd, as $V$ increases, the system will become trivial at the critical point. The behaviors of $M$ agree with the those of $\nu$ we defined above using transfer matrices. 

In summary, we have investigated the properties of the 1D mosaic lattice model with $p$-wave pairing, which constitutes an extension of the 1D Kitaev chain model. We analyze different mosaic lattices corresponding to on-site potentials being periodic, quasiperiodic, or randomly distributed. By checking the energy spectra and the $Z_2$ topological invariant of the system, we find the interplay between mosaic modulation in the on-site potential and superconducting pairing significantly alter the properties of topological superconductor. When the interval of the equally spaced mosaic modulation is odd, the nontrivial phase and Majorana zero modes can only exist within a finite range of the on-site potential except for a few special cases in the commensurate lattices. However, if the interval of the mosaic modulations is even, then the topological phase survives for any finite on-site potential. The robustness of Majorana zero modes in  mosaic lattice is likely to benefit the practical implementation of topological quantum computation.

\begin{acknowledgments}
R. L\"u is supported by NSFC under Grants No. 11874234 and the National Key Research and Development Program of China (2018YFA0306504).
\end{acknowledgments}

\begin{widetext}
\section{Supplementary Material}
%%%%%%%%%% Prefix a "S" to all equations, figures, tables and reset the counter %%%%%%%%%%
\setcounter{equation}{0} \setcounter{figure}{0} \setcounter{table}{0} %
\renewcommand{\theequation}{S\arabic{equation}} \renewcommand{\thefigure}{S%
\arabic{figure}} \renewcommand{\bibnumfmt}[1]{[#1]} \renewcommand{%
\citenumfont}[1]{#1}
%%%%%%%%%% Prefix a "S" to all equations, figures, tables and reset the counter %%%%%%%%%%
In the supplementary material, we will introduce the Bogoliubov-de Gennes transformation and use it to rewrite the Hamiltonian of the one-dimensional (1D) mosaic lattices with $p$-wave pairing, which can be diagonalized conveniently to obtain the system's eigenenergies and the corresponding eigenstates. Then we will present the transfer matrix method to investigate the properties of the 1D topological superconductors and deduce the phase boundaries for different mosaic lattices with commensurate, incommensurate and disordered on-site potentials.

\section{S1. Bogoliubov-de Gennes transformation}
For convenience, we first write down the Hamiltonian of the 1D mosaic lattice with $p$-wave pairing as follows
\begin{equation}\label{H1}
H = \sum_{j=1} [-t c_{j+1}^\dagger c_j + \Delta c_{j+1}^\dagger c_j^\dagger + H.c.] + \sum_{j=1}^{L} V_j c_j^\dagger c_j,
\end{equation}
where $c_j^\dagger$ ($c_j$) is the creation (annihilation) operator at site $j$. $t$ is the hopping amplitude between the nearest neighboring lattice site. $\Delta$ denotes the amplitude of the $p$-wave superconducting pairing in the system. The on-site potential $V_j$ is represented by the following function
\begin{equation}
V_j = \left\lbrace 
\begin{aligned}
& V \cos [ 2 \pi (\alpha j + \phi) ],\qquad j = s \kappa, \\
& 0,\qquad \text{otherwise}.
\end{aligned}
\right. 
\end{equation}
Here, $s=1,2,\cdots,N$. Thus, the on-site potential occurs with interval $\kappa$ and we can define a quasi-cell with the nearest $\kappa$ lattice sites. The system size can be expressed by $L=\kappa N$ with $N$ being the number of the unit cell. Besides, $\alpha$ determines the period of the potential. $\phi$ is a phase factor and we set $\phi=0$ throughout this paper.

To diagonalize the Hamiltonian, we can use the Bogoliubov-de Gennes transformation, which is defined as
\begin{equation}
\eta_n^\dagger = [u_{n,j} c_{n,j}^\dagger + v_{n,j} c_{n,j}],
\end{equation}
where $u_{n,j}$ and $v_{n,j}$ are chosen to be real. Then the wave function of the Hamiltonian is 
\begin{equation}
|\Psi_n \rangle = \eta_n^\dagger |0 \rangle = \sum_{j=1}^{L} [u_{n,j}c_{n,j}^\dagger + v_{n,j} c_{n,j}].
\end{equation}
After substituting the above wave function into the Schr\"odinger equation $H | \Psi \rangle = E_n | \Psi \rangle$ with $E_n$ being the $n$th eigenenergy, we have 
\begin{equation}
\left\{ \begin{aligned}
& -t u_{n,j-1} + \Delta v_{n,j-1} + V_j u_{n,j} - t u_{n,j+1} - \Delta v_{n,j+1} = E_n u_{n,j}, \\
& -\Delta u_{n,j-1} + t v_{n,j-1} - V_j v_{n,j} + \Delta u_{n,j+1} + t v_{n,j+1} = E_n v_{n,j}.
\end{aligned} \right.
\end{equation}
Representing the wave function as a column vector 
\begin{equation}
|\Psi \rangle = [u_{n,1}, v_{n,1}, u_{n,2}, v_{n,2}, \cdots, u_{n,L}, v_{n,L}]^T,
\end{equation}
the model Hamiltonian can be written as a $2L \times 2L$ matrix
\begin{equation}
\mathcal{H}=
  \begin{pmatrix}
    A_1 & B & 0 & \cdots & \cdots & \cdots & C \\
    B^\dagger & A_2 & B & 0 & \cdots & \cdots & 0 \\
    0 & B^\dagger & A_3 & B & 0 & \cdots & 0 \\
    \vdots & \ddots & \ddots & \ddots & \ddots & \ddots & \vdots \\
    0 & \cdots & 0 & B^\dagger & A_{N-2} & B & 0 \\
    0 & \cdots & \cdots & 0 & B^\dagger & A_{N-1} & B \\
    C^\dagger & \cdots & \cdots & \cdots & 0 & B^\dagger & A_N
  \end{pmatrix},
\end{equation}
where
\begin{equation}
  A_j =
  \begin{pmatrix}
    V_j & 0 \\
    0 & -V_j
  \end{pmatrix},
\end{equation}
\begin{equation}
  B=\begin{pmatrix}
      -t & -\Delta \\
      \Delta & t
    \end{pmatrix},
\end{equation}
and
\begin{equation}
  C=\begin{pmatrix}
      -t & \Delta \\
      -\Delta & t.
    \end{pmatrix}
\end{equation}
Here we have assumed periodic boundary condition which implies that $c_{j+L}=c_j$. For the system with open boundary conditions, we can replace the $C$ matrix by a $2 \times 2$ null matrix. Diagonalizing this $\mathcal{H}$ matrix, we can obtain the energy spectrum and eigenstates easily.

\section{S2. Transfer matrix method}
In this section we introduce the transfer matrix method which will be very useful in analyzing the topological properties of the 1D systems. Different from the BdG transformation above, here we decompose the Dirac fermions into Majorana fermions by writing the $c_j$ operator as 
\begin{equation*}
c_j = \frac{1}{2}(a_j + i b_j),
\end{equation*}
where $i$ is the imaginary unit and $a_j$ and $b_j$ are the Majorana fermion operators satisfying the following relations
\begin{equation*}
a_j^2 + b_j^2 = 1, \quad \{a_i, a_j\}=\{b_i, b_j\} =2 \delta_{i,j}, \quad \{a_i,b_j\}=0.
\end{equation*}
Then the system's Hamiltonian can be rewritten in terms of these Majorana fermions as
\begin{equation}
H = \frac{i}{2} \sum_j [(-t + \Delta) a_j b_{j+1} + (t+\Delta)b_j a_{j+1} - V_j a_j b_j].
\end{equation}

Consider Majorana states $Q_A=\sum_j \alpha_j a_j$, $Q_B=\sum_j \beta_j b_j$ ($\alpha_j$ and $\beta_j$ being real), we can define the time-dependent Majorana modes in the Heisenberg representation as $a_j= \alpha_j e^{-i \omega t}$ and $b_j=\beta_j e^{i\omega t}$. Then from the equation of motion we have 
\begin{equation}
\left\{ \begin{aligned}
& (t-\Delta) \alpha_{j-1} + (t+\Delta) \alpha_{j+1} - V_j \alpha _j = -i \omega \beta_j,\\
& -(t+\Delta) \beta_{j-1} - (t-\Delta) \beta_{j+1} + V_j \beta_j = -i \omega \alpha_j.
\end{aligned} \right.
\end{equation}
For the Majorana zero modes, we only need to focus on the $\omega=0$ case and thus, the equations above are decoupled and can be reorganized into the transfer matrix form as
\begin{equation}
\begin{pmatrix} \alpha_{j+1} \\ \alpha_j \end{pmatrix} = A_j \begin{pmatrix} \alpha_j \\ \alpha_{j-1} \end{pmatrix},
\end{equation}
where 
\begin{equation}
A_j = 
\begin{pmatrix}
\frac{V_j}{t+\Delta} & \frac{\Delta-t}{\Delta+t} \\
1 & 0
\end{pmatrix} 
\end{equation}
is the transfer matrix. A similar expression can also be obtained for the $\beta_j$s and the corresponding transfer matrix $B_j$ are related to $A_j$ by $B_j = A_j^{-1}$. The behaviors of the modes at the boundaries of the system are determined by the transfer matrix 
\begin{equation}
\mathcal{A}=\prod_{j=1}^L A_j.
\end{equation}
If the system has periodicity of $p$, then the properties of the modes are determined by 
\begin{equation}
\mathcal{A} = A_p A_{p-1} \cdots A_2 A_1.
\end{equation}

Denote the eigenvalues of $\mathcal{A}$ as $\lambda_1$ and $\lambda_2$, then if $|\lambda_1|, |\lambda_2| < 1$, there will be an a-mode localized at the left end and a b-mode localized at the right end of the 1D lattice. If $|\lambda_1|, |\lambda_2| > 1$, then the two modes will become localized at the opposite ends of the system.

The transfer matrix can be used to define topological invariant, which is very useful in characterizing the topological properties of the 1D system with $p$-wave superconducting pairing. To construct the topological invariant, we can use the characteristic polynomial of $\mathcal{A}$, which is defined as
\begin{equation}
f(z) = \text{det} (\mathcal{A}- Iz).
\end{equation} 
Since $\mathcal{A}$ is a $2\times2$ matrix, the two eigenvalues of the transfer matrix will determine the existence of Majorana bound states at the boundaries. The topological invariant is given as
\begin{equation}
\nu = -\text{sgn}[f(1)f(-1)].
\end{equation}
The index is equal to 1 if and only if there is one eigenvalue of $\mathcal{A}$ whose magnitude is less than 1. The $\nu=1$ indicates the topologically trivial phase. The $\nu=-1$ indicates the nontrivial phase, which reflects that both the eigenvalues of $\mathcal{A}$ are either greater or smaller than 1 in magnitude. The case with $\nu=0$ implies the phase transition where the bulk gap is closed. Next we will use the transfer matrix and the topological invariant $\nu$ to analyze the topological phases and the phase transitions in the 1D mosaic lattices with $p$-wave pairing.

\section{S3. Commensurate mosaic lattice with \texorpdfstring{$\boldsymbol{\kappa=2m}$}{$\kappa=2m$}}
From the energy spectra of the model Hamiltonian in Eq.~(\ref{H1}) with $\kappa$ being an even integer, we find that the bulk energy gap will always being finite and the Majorana zero modes will always exist at the ends of the 1D system, as shown in the main text. Here we prove this by using the transfer matrix method introduced above. For convenience, we set $t$ being positive and $0<\Delta<t$. The situation with $\Delta > 1$ can be analyzed similarly since it is related to the $0<\Delta<t$ case by a proper transformation, as shown in Ref.~\cite{DeGottardi2013PRL}. We further introduce a parameter defined as 
\begin{equation*}
\delta=\frac{\Delta-t}{\Delta+t}. 
\end{equation*}

\subsection{A. Case \texorpdfstring{$\boldsymbol{\kappa=2, \alpha=1/2}$}{$\kappa=2, \alpha=1/2$}}
We first check the case with $\kappa=2$ and $\alpha=1/2$. Then the period of the on-site potential is 2 and we have 
\begin{equation*}
V_j=
\left\{ \begin{aligned}
& 0, \qquad j \quad \text{odd}; \\
& V, \qquad j \quad \text{even}.
\end{aligned} \right.
\end{equation*}
According to the transfer matrix method we have introduced above, the properties of the boundary modes are determined by the transfer matrix, which in this case is 
\begin{equation}
\mathcal{A}=A_1 A_2=\begin{pmatrix} 0 & \delta \\ 1 & 0 \end{pmatrix}
\begin{pmatrix}
\frac{V}{\Delta+t} & \delta \\ 1 & 0 \end{pmatrix}
=\begin{pmatrix}
\delta & 0 \\ \frac{V}{\Delta+t} & \delta
\end{pmatrix}.
\end{equation}
Thus, the characteristic polynomial of $\mathcal{A}$ is $f(z)=\text{det}(\mathcal{A}-Iz)=(\delta-z)^2$ and we have
\begin{align}
f(1)f(-1) = (\delta-1)^2 (\delta+1)^2 = (\delta^2-1)^2 > 0.
\end{align}
So $\nu = -\text{sgn}[f(1)f(-1)] = -1$ is always established irrespective of the value of $V$. The system is always in the topologically nontrivial phase, which is consistent with the results we observed from the energy spectrum.

\subsection{B. Case \texorpdfstring{$\boldsymbol{\kappa=4, \alpha=1/2}$}{$\kappa=4, \alpha=1/2$}}
We further check the case with $\kappa=4$, $\alpha=1/2$. Now the period of the on-site potential is $4$ and the transfer matrices are 
\begin{equation}
A_1=A_2=A_3=\begin{pmatrix} 0 & \delta \\ 1 & 0 \end{pmatrix}, \quad A_4=\begin{pmatrix}
\frac{V}{\Delta+t} & \delta \\ 1 & 0 \end{pmatrix}, \quad
\mathcal{A}=A_1 A_2 A_3 A_4 = \begin{pmatrix}
\delta^2 & 0 \\ \frac{V \delta}{\Delta + t} & \delta^2
\end{pmatrix}.
\end{equation}
Then the characteristic polynomial of $\mathcal{A}$ is $f(z)=(\delta^2-z)^2$ and $f(1)f(-1)=(\delta^4-1)^2$. So the topological invariant $\nu=-\text{sgn}[f(1)f(-1)]=-1$. Again, the system is always in the topological phase, no matter how strong the on-site potential will be.

\subsection{C. General case \texorpdfstring{$\boldsymbol{\kappa=2m}$}{$\kappa=2m$}}
Suppose the length of the lattice is $L=\kappa N$ with $N$ being a large positive integer. The on-site potential will be nonzero only at the ($s \kappa$)th site with $s=1,2,\cdots,N$. The corresponding transfer matrix for these sites are of the following form
\begin{equation*}
A_{s \kappa} = \begin{pmatrix}
\frac{V_{s \kappa}}{\Delta + t} & \delta \\ 1 & 0
\end{pmatrix}.
\end{equation*}
On the other hand, the transfer matrices for the sites with zero on-site potential are the same, i.e.
\begin{equation*}
A_{j \neq s \kappa} = \begin{pmatrix}
0 & \delta \\ 1 & 0
\end{pmatrix}.
\end{equation*}
In the mosaic lattice, the matrix $A_{s \kappa}$ appears every $\kappa$ sites when calculating the transfer matrix $\mathcal{A}=\prod_{j=1}^{L} A_j$. We can decompose the product into $N$ parts, each of which includes $\kappa$ sites. In each part, there will be ($\kappa-1$) transfer matrices that are of the same form, i.e., $A_{j \neq s \kappa}$, their product can be calculated easily as 
\begin{equation}\label{Ak}
A_1 A_2 \cdots A_{\kappa-1} = \left\{ \begin{aligned} 
\begin{pmatrix} 0 & \delta^m \\ \delta^{m-1} & 0 \end{pmatrix}, \qquad \kappa=2m; \\
\begin{pmatrix} \delta^m & 0 \\ 0 & \delta^m \end{pmatrix}, \qquad \kappa=(2m+1);
\end{aligned} \right.
\end{equation}
with $m=1,2,\cdots$.

Now we turn to the general cases where the parameter $\kappa$ is an even number and we set $\kappa=2m$. The lattice can be either commensurate or incommensurate. Then the transfer matrix of the system is 
\begin{equation}
\begin{aligned}
\mathcal{A}&= A_1 A_2 A_3 \cdots A_{\kappa N} \\
&= (A_1 A_2 \cdots A_\kappa)(A_{\kappa+1} A_{\kappa+2} \cdots A_{2\kappa}) \cdots (A_{(N-1)\kappa+1} A_{(N-1)\kappa+2} \cdots A_{N \kappa})\\
&= \begin{pmatrix} \delta^m & 0 \\ \frac{V_{m}\delta^{m-1}}{\Delta+t} & \delta^{m} \end{pmatrix}
\begin{pmatrix} \delta^m & 0 \\ \frac{V_{4m}\delta^{m-1}}{\Delta+t} & \delta^{m} \end{pmatrix} \cdots
\begin{pmatrix} \delta^m & 0 \\ \frac{V_{2mN}\delta^{m-1}}{\Delta+t} & \delta^{m} \end{pmatrix} \\
&= \begin{pmatrix} \delta^{mN} & 0 \\ \frac{(\sum_{j=1}^{\kappa N}V_{j})\delta^{mN-1}}{\Delta+t} & \delta^{mN} \end{pmatrix}.
\end{aligned}
\end{equation}
Obviously, the transfer matrix for the whole lattice is of the form $\begin{pmatrix}
a & 0 \\ b & a \end{pmatrix}$. We can calculate the characteristic polynomial of $\mathcal{A}$ in this case 
\begin{equation}
f(z) = \text{det}(\mathcal{A}-Iz)=(\delta^{mN}-z)^2.
\end{equation}
and we have $f(1)f(-1)=(\delta^{2m N}-1)^2 > 0$. The topological invariant $\nu=-\text{sgn}[f(1)f(-1)]=-1$ and does not depend on the value of $V$. Thus we prove that as long as $\kappa$ is an even integer, the system is always in the topologically nontrivial phase.

\section{S3. Commensurate mosaic lattice with \texorpdfstring{$\boldsymbol{\kappa=(2m+1)}$}{$\kappa=(2m+1)$}}
Here we first analyze specific examples with $\kappa=1$ and $3$. Then we will discuss the general cases.

\subsection{A. Case \texorpdfstring{$\boldsymbol{\kappa=1, \alpha=1/2}$}{$\kappa=1, \alpha=1/2$}}
If we set $\kappa=1$ and $\alpha=1/2$, then the period of the on-site potential is $2$. The on-site potentials in each unit cell are denoted as $V_1=-V$ and $V_2=V$. So the transfer matrix of the system is 
\begin{equation}
\mathcal{A}=A_1 A_2 = \begin{pmatrix} \frac{-V}{\Delta+t} & \delta \\ 1 & 0 \end{pmatrix}
\begin{pmatrix} \frac{V}{\Delta+t} & \delta \\ 1 & 0 \end{pmatrix}
=\begin{pmatrix} \frac{-V^2}{(\Delta+t)^2}+\delta & \frac{-V\delta}{\Delta+t} \\ \frac{V}{\Delta+t} & \delta \end{pmatrix}.
\end{equation}
Then the characteristic polynomial of $\mathcal{A}$ is
\begin{equation}
f(z)=\text{det}(\mathcal{A}-Iz)=(\delta-z)^2 + \frac{V^2 z}{(\Delta+t)^2}. 
\end{equation}
Thus the topological invariant can be calculated as
\begin{equation}
\nu=\text{sgn}[f(1)f(-1)]=\text{sgn} \left\{ \left[ (\delta-1)^2 + \frac{V^2}{(\Delta+t)^2} \right] \left[ (\delta+1)^2 - \frac{V^2}{(\Delta+t)^2} \right ] \right\}.
\end{equation}
The index $\nu$ depends on the value of $V$ and there will be phase transitions when we tune the on-site potential. From $\nu=0$ we can deduce the phase transition point $V_c$, which gives us 
\begin{equation}
(\delta+1)^2-\frac{V_c^2}{(\Delta+t)^2}=0 \rightarrow V_c = \pm[(\delta+1)(\Delta+t)].
\end{equation}
Take $t=1$ and $\Delta=0.5$ as in the main text, we have $V_c=\pm1$, which agrees with the numerical results we obtain from the energy spectrum of the system.

\subsection{B. Case \texorpdfstring{$\boldsymbol{\kappa=3, \alpha=1/2}$}{$\kappa=3, \alpha=1/2$}}
For the mosaic lattice with $\kappa=3, \alpha=1/2$, the period of the system is $6$. We denote the on-site potential as $V_j$ with $j=1,2,\cdots,6$ in each unit cell. There are only two sites with nonzero on-site potential, i.e. $V_3 = -V$ and $V_6=V$. While $V_1=V_2=V_4=V_5=0$. The transfer matrices for these sites are 
\begin{equation*}
A_1=A_2=A_4=A_5=\begin{pmatrix} 0 & \delta \\ 1 & 0 \end{pmatrix}; \quad
A_3=\begin{pmatrix} \frac{-V}{\Delta+t} & \delta \\ 1 & 0 \end{pmatrix}; \quad
A_6=\begin{pmatrix} \frac{V}{\Delta+t} & \delta \\ 1 & 0 \end{pmatrix}.
\end{equation*}  
So we have 
\begin{equation}
\mathcal{A}=A_1 A_2 A_3 A_4 A_5 A_6 = \delta^2 \begin{pmatrix}
\frac{-V^2}{(\delta+t)^2}+\delta & \frac{-V}{\Delta+t} \\ \frac{V}{\Delta+t} & \delta
\end{pmatrix}.
\end{equation}
The characteristic polynomial of this transfer matrix $\mathcal{A}$ is 
\begin{equation}
f(z) = \text{det}(\mathcal{A}-Iz)=\left[ -\frac{V^2 \delta^2}{(\Delta+t)^2}+\delta^3 - z \right] \left[ \delta^3-z \right] + \frac{V^2 \delta^5}{(\Delta+t)^2}.
\end{equation}
Then we can obtain
\begin{equation}
f(1)=\delta^6-2\delta^3+1+\frac{V^2 \delta^2}{(\Delta+t)^2}, \qquad f(-1)=\delta^6+2\delta^3+1-\frac{V^2 \delta^2}{(\Delta+t)^2}.
\end{equation}
Here we see that the topological invariant $\nu=-\text{sgn}[f(1)f(-1)]$ depends on the value of $V$, which means that as we tune the strength of the on-site potentials, the system will undergo a phase transition from topologically nontrivial phase to the trivial phase or vice verse. The index $\nu=0$ indicates the critical point of the phase transition. By solving the following equation
\begin{equation*}
f(1)f(-1)=(\delta^6+1)^2-\delta^4 [ \frac{V_c^2}{(\Delta+t)^2}-2\delta ]^2 = 0,
\end{equation*}
we get the critical strength of the on-site potential as
\begin{equation}
V_c = \pm (\Delta + t) \sqrt{\delta^4+\frac{1}{\delta^2}+2\delta}.
\end{equation}
Take $t=1$ and $\Delta=0.5$, we have $V_c=\pm \frac{13}{3} \approx 4.33$, which agrees with the bulk gap closing point in the energy spectrum shown in the main text.

\subsection{C. Case \texorpdfstring{$\boldsymbol{\kappa=3, \alpha=1/4}$}{$\kappa=3, \alpha=1/4$}}
In the lattice with $\kappa=3$ and $\alpha=1/4$, we can show the system is always topologically nontrivial. The on-site potential becomes $V_j=V \cos(2\pi \alpha j)=V \cos(\frac{j\pi}{2})$ for $j=s \kappa$. The period of the system is $12$, we can label the sites in each cell as $1,2,\cdots,12$. Only the $s\kappa$th will have nonzero on-site potential. However, since $\alpha=1/4$, we have $V_3=V_9=0$ and $V_6=-V_{12}=-V$. Splitting the cell into four parts each of which contains three sites, then the transfer matrix for each part is
\begin{equation*}
A_1 A_2 A_3 =A_7 A_8 A_9 = \begin{pmatrix} 0 & \delta^2 \\ \delta & 0 \end{pmatrix};\quad
A_4 A_5 A_6 = \begin{pmatrix} \frac{-V}{\Delta+t} & \delta^2 \\ \delta & 0 \end{pmatrix}; \quad
A_{10} A_{11} A_{12} = \begin{pmatrix} \frac{V}{\Delta+t} & \delta^2 \\ \delta & 0 \end{pmatrix}.
\end{equation*} 
Then the transfer matrix of the whole unit cell becomes 
\begin{equation}
\mathcal{A}=A_1 A_2 \cdots A_{12} = \begin{pmatrix} \delta^6 & 0 \\ 0 & \delta^6 \end{pmatrix}.
\end{equation}
So the characterize polynomial and the topological invariant $\nu$ are independent of the on-site potential and we always have $\nu=-1$. The system is always topologically nontrivial in these cases, even though $\kappa$ is odd. 

\subsection{D. General case \texorpdfstring{$\boldsymbol{\kappa=(2m+1)}$}{$\kappa=(2m+1)$}}
In the mosaic lattices with an odd $\kappa$, we split the whole lattice into $N$ unit cell (or quasi-cell) each of which contains $\kappa$ sites and only the $\kappa$th site has nonzero on-site potential. From Eq.~(\ref{Ak}), we know that the product of the ($\kappa-1$) transfer matrix for the site with zero on-site potential in each cell will be a diagonal matrix. So the transfer matrix for the whole cell is 
\begin{equation}
\mathcal{A}=(A_{s1} A_{s2} \cdots A_{s,2m}) A_{s,2m+1} = \begin{pmatrix} \delta^m & 0 \\ 0 & \delta^m \end{pmatrix} A_{s,2m+1}
=\begin{pmatrix} \delta^m & 0 \\ 0 & \delta^m \end{pmatrix} \begin{pmatrix} \frac{V_{j=s \kappa}}{\Delta+t} & \delta \\ 1 & 0 \end{pmatrix} = \begin{pmatrix} \frac{V_{j=s \kappa} \delta^m}{\Delta+t} & \delta^{m+1} \\ \delta^m & 0 \end{pmatrix}.
\end{equation} 
Different from the cases with $\kappa$ being even, here the transfer matrix $\mathcal{A}$ is not defective and we see that the on-site potential appears in the diagonal term, which means that the characteristic polynomial $f(z)$ and the corresponding topological invariant $\nu$ will depend on the value of $V$. Thus the topological phase in the systems with odd $\kappa$ will be determined by the on-site terms. 

However, if $q=2^k$ ($k \geq 2$) with $k$ being integers in the commensurate lattice $\alpha=p/q$, the situation becomes different. In this case, the on-site potential is $V_j = V\cos(2\pi \alpha j)=V \cos(\frac{pj\pi}{2^{k-1}})$. So apart from those sites with $j \neq s \kappa$, the on-site potential at site $j=2^{k-2}\kappa$ is also zero. Again we take $\alpha=1/4$ as an example. There are $4\kappa$ sites in each unit cell and we label these site in one unit cell as $j=1,2,\cdots,4\kappa$. According to the mosaic form of the on-site potentials, we have
\begin{equation*}
V_\kappa=0,\quad V_{2\kappa}=-V,\quad V_{3\kappa}=0,\quad V_{4\kappa}=V,
\end{equation*}
and all the potentials at the remaining sites in the cell are zero. Since $\kappa=(2m+1)$ is an odd number, the number of sites between the $s\kappa$th and $(s+1)\kappa$th site is $2m$, and the transfer matrices are of the same form, i.e. $\begin{pmatrix} 0 & \delta \\ 1 & 0 \end{pmatrix}$. Their product gives
\begin{equation*}
A_1 A_2 \cdots A_{2m} = \begin{pmatrix} \delta^m & 0 \\ 0 & \delta^m \end{pmatrix},
\end{equation*}
which is diagonal. Then the transfer matrix of the whole unit cell is 
\begin{align*}
\mathcal{A}&=A_1 A_2 \cdots A_{4\kappa}=\begin{pmatrix} \delta^{4m} & 0 \\ 0 & \delta^{4m} \end{pmatrix} A_\kappa A_{2\kappa} A_{3\kappa} A_{4\kappa} \\
&=\begin{pmatrix} \delta^{4m} & 0 \\ 0 & \delta^{4m} \end{pmatrix} \begin{pmatrix} \frac{V_\kappa}{\Delta+t} & \delta \\ 1 & 0 \end{pmatrix} \begin{pmatrix} \frac{V_{2\kappa}}{\Delta+t} & \delta \\ 1 & 0 \end{pmatrix} \begin{pmatrix} \frac{V_{3\kappa}}{\Delta+t} & \delta \\ 1 & 0 \end{pmatrix} \begin{pmatrix} \frac{V_{4\kappa}}{\Delta+t} & \delta \\ 1 & 0 \end{pmatrix}= \begin{pmatrix} \delta^{4m+2} & 0 \\ 0 & \delta^{4m+2} \end{pmatrix}.
\end{align*}
So in this case, even though $\kappa$ is odd, the transfer matrix of the unit cell is independent of the on-site terms and the topological invariant $\nu=-1$ for any values of $V$. The system will always be in the nontrivial phase no matter how strong $V$ is. We can prove similarly for other cases with $q=2^{k}$ ($k \geq 2$).

For the commensurate lattices, we can use the transfer matrix method to calculate the critical value conveniently. However, if the lattice is incommensurate, the situation becomes more complicated since we have to take all the sites with nonzero on-site potentials into consideration. For the topological superconductor in the incommensurate lattices, it is well known that the topological phase will be destroyed by the Anderson localization phase transition as the quasiperiodic on-site potential becomes stronger than the critical value~\cite{Cai2013PRL}. Here in the incommensurate mosaic lattice, we also observe such phenomenon when $\kappa$ is odd from the energy spectra. To obtain the critical values for the phase transition, we can use the Lyapunov exponent~\cite{DeGottardi2013PRL}, which will be discussed in the next section. 

\section{S4. Incommensurate mosaic lattices}
The transfer matrix method can be used to determine the Anderson localization properties of the normal disordered system~\cite{Motrunich2001PRB}. As to the model we study in this paper, we can use similar method to study the topological phase transition in the quasiperiodic latices~\cite{DeGottardi2013PRL}. To connect the model with $p$-wave superconducting pairing with the normal lattice, we can perform a similarity transformation to the transfer matrix as follows
\begin{equation}
A_j = \sqrt{\delta'} S \tilde{A}_j S^{-1},
\end{equation}
where $S=\text{diag}(\delta'^{1/4},1/\delta'^{1/4})$ and $\delta'=-\delta=\frac{t-\Delta}{\Delta+t}$. Notice that we have set $0<\Delta<t$. The transfer matrix $\tilde{A}_j$ is
\begin{equation}
\tilde{A}_j = \begin{pmatrix}
\frac{V_j}{\sqrt{t^2-\Delta^2}} & -1 \\ 1 & 0 \end{pmatrix}.
\end{equation}
Then we have 
\begin{equation}\label{At}
\mathcal{A}(V,\Delta)=\left( \sqrt{\frac{t-\Delta}{t+\Delta}} \right)^L S \mathcal{A}\left( \frac{V}{\sqrt{t^2-\Delta^2}},0 \right) S^{-1}.
\end{equation}
We have already introduced the topological invariant $\nu=\text{sgn}[f(1)f(-1)]$ with $f(z)=\text{det}(\mathcal{A}-Iz)$ being the characteristic polynomial of $\mathcal{A}$. Since we take $\Delta$ to be positive, we have $|\text{det}\mathcal{A}|<1$. The two eigenvalues of $\mathcal{A}$, $\lambda_1$ and $\lambda_2$, obey the relation $|\lambda_1 \lambda_2|<1$. Therefore, if we take $\lambda_1$ to be the smaller eigenvalue in magnitude, namely $|\lambda_1|<|\lambda_2|$, then we have $|\lambda_1|<1$. Then the topological invariant $\nu$ if totally determined by $\lambda_2$ and we have $\nu=\text{sgn}(\text{ln} |\lambda_2|)$. $|\lambda_2|=1$ gives the critical point of the phase transition.

By taking the logarithm of the eigenvalues of Eq.~(\ref{At}) and then imposing the condition $|\lambda_2|=1$ lead to
\begin{equation}\label{Ly}
\gamma(V,\Delta)=\gamma \left( \frac{V}{\sqrt{t^2-\Delta^2}}, 0 \right) - \frac{1}{2} \text{ln} \left( \frac{t+\Delta}{t-\Delta} \right).
\end{equation}
Here we have defined the Lyapunov exponent as 
\begin{equation}
\gamma(V,\Delta) \equiv \lim_{L \rightarrow \infty} \frac{1}{L} \text{ln} |\lambda_2(V,\Delta)|.
\end{equation}
The Lyapunov exponent is related to the localization length $l$ by $\gamma=1/l$. In the limit $\gamma(V,\Delta) \rightarrow 0$, Eq.~(\ref{Ly}) gives the phase boundary between the topologically nontrivial and trivial phase. For instance, in the incommensurate lattice $V_j=V\cos(2\pi \alpha j)$ with $\alpha$ being an irrational number, if there is no $p$-wave pairing in the system, the normal state Lyapunov exponent $\gamma(V,0)=\ln (V/2)$ for $V>2$ and $0$ for $0<V<2$. Then from Eq.~(\ref{Ly}) predicts the topologically nontrivial phase with the presence of $p$-wave pairing when $|V|>2(t+\Delta)$.

For the mosaic lattices we consider in this paper, it is known that in the normal state with $\Delta=0$, there will be mobility edges in the energy spectra, as shown in Ref.~\cite{Wang2020arxiv}. It seems that we cannot use the above method directly to the mosaic lattices with $p$-wave pairing. Since the characteristic feature of the topological phase is the existence of zero-energy modes, we can focus on the states with zero-energy. Following the discussions in Ref.~\cite{Wang2020arxiv}, if the energy $E$ lies in the spectrum, the Lyapunov exponent of the 1D mosaic lattice with $\Delta=0$ is given by $\kappa \gamma_0(E)=\text{max} \{\ln |Va_\kappa/2|,0 \}$. When $|Va_\kappa/2|>1$, the Lyapunov exponent is 
\begin{equation}
\gamma_0(E) = \frac{\ln |\frac{V}{2} a_\kappa|}{\kappa}.
\end{equation}
Notice that we have $V=2\lambda$ comparing with the model in Ref.~\cite{Wang2020arxiv}. Besides, $a_\kappa$ is given by
\begin{equation}
a_\kappa=\frac{1}{\sqrt{E^2-4}} \left[ \left( \frac{E+\sqrt{E^2-4}}{2} \right)^\kappa - \left( \frac{E-\sqrt{E^2-4}}{2} \right)^\kappa \right].
\end{equation}
If we set $E=0$, we have 
\begin{equation}
a_\kappa = \frac{1}{2i} \left[ i^\kappa - (-i)^\kappa \right].
\end{equation}
Then the Lyapunov exponent for the states with zero energy is 
\begin{equation}
\gamma_0(0)= \left\{ \begin{aligned} 
& 0, \qquad \kappa \quad \text{even};\\
& \frac{\ln \frac{|V|}{2}}{\kappa}, \qquad \kappa \quad \text{odd}.
\end{aligned} \right.
\end{equation}
Thus, when $\kappa$ is an even integer, the Lyapunov exponent at zero-energy will always be zero and the corresponding states will always being delocalized. So in the mosaic lattice with $p$-wave pairing, the zero-energy modes will always present and will not be localized by the quasiperiodic potentials when $\kappa$ is even. The system is topologically nontrivial no matter how strong the on-site potentials will be. However, for the system with $\kappa$ being odd, the situation is totally different. As the incommensurate potential becomes stronger, the zero-energy states will be localized, so there will be a phase transition in the system with $p$-wave pairing. The phase boundary can be deduced by using Eq.~(\ref{Ly}), which leads to 
\begin{equation}
|V_c|=2 \sqrt{\frac{(t+\Delta)^{\kappa+1}}{(t-\Delta)^{\kappa-1}}} \xlongequal{\kappa=2m+1} \frac{2(t+\Delta)^{m+1}}{(t-\Delta)^m}, \qquad (m=0,1,2,\cdots).
\end{equation}
Taking $t=1$ and $\Delta=0.5$, we get $|V_c|=9$ for the incommensurate mosaic lattice with $\kappa=3$, which agrees with the gap closing point in the energy spectrum.

From all the discussions above, we can conclude that both in the commensurate and incommensurate mosaic lattices with $p$-wave pairing, the system will always in the topologically nontrivial phase as long as the parameter $\kappa$ is even. The Majorana zero modes at the ends of the 1D systems thus become very robust. On the contrary, when $\kappa$ is odd, there will be a topological phase transition where the system will enters into the trivial phase as the on-site potential becomes stronger than the critical values and the Majorana zero modes will disappear. 

\section{S5. Disordered mosaic lattices with \texorpdfstring{$\boldsymbol{p}$}{$p$}-wave pairing}
We have investigate the 1D mosaic lattices with periodic or quasiperiodic on-site potentials and found that the parameter $\kappa$ can influence the topological properties of the 1D topological superconducting system significantly. If the on-site potentials in the mosaic lattices have no specific forms but are random, then what will happen? Consider the 1D model in Eq.~(\ref{H1}) with the following on-site potentials
\begin{equation}
V_j = \left\lbrace 
\begin{aligned}
& w_j \in [-W,W],\qquad j = s \kappa, \\
& 0,\qquad \text{otherwise}.
\end{aligned}
\right. 
\end{equation}
Here the nonzero on-site potentials at the $s\kappa$th are uniformly distributed in $[-W,W]$ with $W$ being and positive number. It is known that for a normal 1D lattice with disorders, all the eigenstates will be localized. The interesting part for the model we study here is that we can prove that when $\kappa$ is even, the system will always be nontrivial without respective to the disordered on-site potentials. The proof is given as follows. 

We first decompose the lattice into $N$ unit with each unit cell containing $\kappa$ sites. In each unit cell, the on-site potentials of all the sites except the $s\kappa$th one are zero. According to Eq.~(\ref{Ak}), when $\kappa=2m$, the product of the transfer matrix in the $s$th unit cell is
\begin{equation}
A_{s1} A_{s2} \cdots A_{s\kappa} = \begin{pmatrix} 0 & \delta^m \\ \delta^{m-1} & 0 \end{pmatrix}
\begin{pmatrix} \frac{V_{s\kappa}}{\Delta+t} & \delta \\ 1 & 0 \end{pmatrix}
=\begin{pmatrix} \delta^m & 0 \\ \frac{V_{s\kappa}}{\Delta+t} & \delta^m \end{pmatrix}.
\end{equation}
Then the transfer matrix for the whole lattice can be obtained by multiplying all these matrices above for unit cells, which leads to 
\begin{equation}
\mathcal{A}=\begin{pmatrix}
\delta^{mN} & 0 \\ \frac{(\sum_{j=1}^L V_j) \delta^{mN-1}}{\Delta+t} & \delta^{mN}
\end{pmatrix}.
\end{equation}
Again, the matrix is defective and there is no on-site potential present in the diagonal terms. The characteristic polynomial of $\mathcal{A}$ is $f(z)=\text{det}[\mathcal{A}-Iz]=(\delta^{mN}-z)^2$, so the topological invariant of the system is 
\begin{equation}
\nu = -\text{sgn}[f(1)f(-1)] = -\text{sgn}\left[ (\delta^{2mN}-1)^2 \right] = -1.
\end{equation}
Hence, the topological invariant is independent of the on-site potentials. The system is always in the topologically nontrivial phase. We can also see that the topological properties do not depend on the specific form of the on-site terms. As long as $\kappa$ is even, we can get the nontrivial phase and observe the Majorana zero modes at the ends of the 1D lattices. 

On the other hand, if $\kappa$ is odd, then the topological phase will be destroyed by the disorders, which is similar to the periodic and quasiperiodic systems we discussed in the above sections. 
\end{widetext}


\begin{thebibliography}{}
\bibitem{Alicea2012PRP}{J. Alicea, Rep. Prog. Phys. \textbf{75,} 076501 (2012).}

\bibitem{Beenakker2013ARCMP}{C. W. J. Beenakker, Annu. Rev. Condens. Matter Phys. \textbf{4,} 113 (2013).}

\bibitem{Elliott2015RMP}{S. R. Elliott and M. Franz, Rev. Mod. Phys. \textbf{87,} 137 (2015).}

\bibitem{Ando2015ARCMP}{Y. Ando and L. Fu, Annu. Rev. Condens. Matter Phys. \textbf{6,} 361 (2015).}

\bibitem{Sato2017RPP}{M. Sato and Y. Ando, Rep. Prog. Phys. \textbf{80,} 076501 (2017).}

%Majorana zero modes and topological quantum computation
\bibitem{Sarma2015npj}{S. Das Sarma, M. Freedman, and C. Nayak, npj Quantum Inf \textbf{1,} 15001 (2015).}

%Majorana-Based Fermionic Quantum Computation
\bibitem{Obrien2018PRL}{T. E. O'Brien, P. Ro\.zek, and A. R. Akhmerov, Phys. Rev. Lett. \textbf{120,} 220504 (2018).}

%Topological quantum computation based on chiral Majorana fermions
\bibitem{Lian2018PNAS}{ B. Lian, X.-Q. Sun, A. Vaezi, X.-L. Qi, and S.-C. Zhang, Proc. Natl. Acad. Sci. U.S.A. \textbf{115,} 10938 (2018).}

%Quantum computing with Majorana fermion codes
\bibitem{Litinski2018PRB}{D. Litinski and F. von Oppen, Phys. Rev. B \textbf{97,} 205404 (2018).}

%theoretical proposal for the realization of MZMs in different platforms
%1D p-wave superconductors
\bibitem{Lutchyn2010PRL}{R. M. Lutchyn, J. D. Sau, and S. Das Sarma, Phys. Rev. Lett. \textbf{105,} 077001 (2010).}

\bibitem{Oreg2010PRL}{Y. Oreg, G. Refael, and F. von Oppen, Phys. Rev. Lett. \textbf{105,} 177002 (2010).}

%2D p+ip superconductors
\bibitem{Stone2004PRB}{M. Stone and R. Roy, Phys. Rev. B \textbf{69,} 184511 (2004).}

\bibitem{Fendley2007PRB}{P. Fendley, M. P. A. Fisher, and C. Nayak, Phys. Rev. B \textbf{75,} 045317 (2007).}

\bibitem{Raghu2010PRL}{S. Raghu, A. Kapitulnik, and S. A. Kivelson, Phys. Rev. Lett. \textbf{105,} 136401 (2010).}

\bibitem{Chung2011PRB}{S. B. Chung, H. J. Zhang, X. L. Qi, and S. C. Zhang, Phys. Rev. B \textbf{84,} 060510(R) (2011).}

%superfluid He-3 
\bibitem{Kopnin1991PRB}{N. B. Kopnin and M. M. Salomaa, Phys. Rev. B \textbf{44,} 9667 (1991).}

\bibitem{Qi2009PRL}{X. L. Qi, T. L. Hughes, S. Raghu, and S. C. Zhang, Phys. Rev. Lett. \textbf{102,} 187001 (2009).}

\bibitem{Chung2009PRL}{S. B. Chung and S. C. Zhang, Phys. Rev. Lett. \textbf{103,} 235301 (2009).}

%ultracold atoms 
\bibitem{Liu2012PRA}{X. J. Liu, L. Jiang, H. Pu, and H. Hu, Phys. Rev. A \textbf{85,} 021603(R) (2012).}

\bibitem{Qu2013NatCom}{C. Qu, Z. Zheng, M. Gong, Y. Xu, L. Mao, X. Zou, G. Guo, and C. Zhang, Nat. Commun. \textbf{4,} 2710 (2013).}

\bibitem{Chen2013PRL}{C. Chen, Phys. Rev. Lett. \textbf{111,} 235302 (2013).}

\bibitem{Ruhman2015PRL}{J. Ruhman, E. Berg, and E. Altman, Phys. Rev. Lett. \textbf{114,} 100401 (2015).}

%magnetic atoms 
\bibitem{Perge2013PRB}{S. Nadj-Perge, I. K. Drozdov, B. A. Bernevig, and A. Yazdani, Phys. Rev. B \textbf{88,} 020407(R) (2013).}

\bibitem{Hui2015SR}{H. Y. Hui, P. M. R. Brydon, J. D. Sau, S. Tewari, and S. Das Sarma, Sci. Rep. 5, 8880 (2015).}

\bibitem{Dumitrescu2015PRB}{E. Dumitrescu, B. Roberts, S. Tewari, J. D. Sau, and S. Das Sarma, Phys. Rev. B \textbf{91,} 094505 (2015).}

%theoretical proposals for the detection of MZMs in 1D TSCs
\bibitem{Ioselevich2011PRL}{P. A. Ioselevich and M. V. Feigel’man Phys. Rev. Lett. \textbf{106,} 077003 (2011).}

\bibitem{Zazunov2011PRB}{A. Zazunov, A. L. Yeyati, and R. Egger, Phys. Rev. B 84, 165440 (2011).}

\bibitem{Wu2012PRB}{B. H. Wu and J. C. Cao, Phys. Rev. B 85, 085415 (2012).}

\bibitem{Jose2012PRL}{Pablo San-Jose, Elsa Prada, and Ramón Aguado, Phys. Rev. Lett. \textbf{108,} 257001 (2012).}

\bibitem{Ueda2014PRB}{A. Ueda and T. Yokoyama, Phys. Rev. B 90, 081405(R) (2014).}

\bibitem{Zeng2016FP}{Q.-B. Zeng, S. Chen, L. You, and R. L\"u, Front. Phys. \textbf{12,} 127302 (2016).}

%experimental detection of Majorana zero modes in 1D TSCs
\bibitem{Mourik2012Science}{V. Mourik, K. Zuo, S. M. Frolov, S. R. Plissard, E. P. A. M. Bakkers, and L. P. Kouwenhoven, Science \textbf{336,} 1003 (2012).}

\bibitem{Das2012NatPhys}{A. Das, Y. Ronen, Y. Most, Y. Oreg, M. Heiblum, and H. Shtrikman, Nat. Phys. 8, 887 (2012).}

\bibitem{Finck2013PRL}{A. D. K. Finck, D. J. Van Harlingen, P. K. Mohseni, K. Jung, and X. Li, Phys. Rev. Lett. \textbf{110,} 126406 (2013).}

\bibitem{Perge2014Science}{S. Nadj-Perge, I. K. Drozdov, J. Li, H. Chen, S. Jeon, J. Seo, A. H. MacDonald, B. A. Bernevig, and A. Yazdani, Science \textbf{346,} 602 (2014).}

\bibitem{Kitaev2001}{A. Y. Kitaev, Phys. Usp. \textbf{44,} 131 (2001).}

\bibitem{Anderson1958PR}{P. W. Anderson, Phys. Rev. \textbf{109,} 1492 (1958).}

\bibitem{Motrunich2001PRB}{O. Motrunich, K. Damle, and D. A. Huse, Phys. Rev. B \textbf{63,} 224204 (2001).}

\bibitem{Potter2010PRL}{A. C. Potter and P. A. Lee, Phys. Rev. Lett. \textbf{105,} 227003 (2010).}

\bibitem{Shivamoggi2010PRB}{V. Shivamoggi, G. Refael, and J. E. Moore. Phys. Rev. B \textbf{82,} 041405(R) (2010).}

\bibitem{Akhmerov2011PRL}{A. R. Akhmerov, J. P. Dahlhaus, F. Hassler, M. Wimmer, and C. W. J. Beenakker, Phys. Rev. Lett. \textbf{106,} 057001 (2011).}

\bibitem{Brouwer2011PRL}{P.W. Brouwer, M. Duckheim, A. Romito, and F. von Oppen, Phys. Rev. Lett. \textbf{107,} 196804 (2011).}

\bibitem{Lobos2012PRL}{A. M. Lobos, R. M. Lutchyn, and S. Das Sarma, Phys. Rev. Lett. \textbf{109,} 146403 (2012).}

\bibitem{Sau2012NC}{J. D. Sau and S. Das Sarma, Nat. Commun. \textbf{3,} 964 (2012).}

\bibitem{Adagideli2014PRB}{\.I. Adagideli, M. Wimmer, and A. Teker, Phys. Rev. B \textbf{89,} 144506 (2014).}

%Topological Superconductor to Anderson Localization Transition in One-Dimensional Incommensurate Lattices
\bibitem{Cai2013PRL}{X. Cai, L.-J. Lang, S. Chen, and Y. Wang, Phys. Rev. Lett. \textbf{110,} 176403 (2013).}

%Majorana Fermions in Superconducting 1D Systems Having Periodic, Quasiperiodic, and Disordered Potentials
\bibitem{DeGottardi2013PRL}{W. DeGottardi, D. Sen, and S. Vishveshwara, Phys. Rev. Lett. \textbf{110,} 146404 (2013).}

%Generalized Aubry-Andre-Harper model with p-wave superconducting pairing
\bibitem{Zeng2016PRB}{Q.-B. Zeng, R. L\"u, and S. Chen, Phys. Rev. B, \textbf{94,} 125408 (2016).}

%One dimensional quasiperiodic mosaic lattice with exact mobility edges
\bibitem{Wang2020arxiv}{Y. Wang, X. Xia, L. Zhang, H. Yao, S. Chen, J. You, Q. Zhou, and X.-J. Liu, Phys. Rev. Lett. \textbf{125,} 196604 (2020).}

\bibitem{SM}{Supplementary material.}

\bibitem{DeGottardi2011NJP}{W. DeGottardi, D. Sen, and S. Vishveshwara, New J. Phys. \textbf{13,} 065028 (2011).}

\end{thebibliography}
\end{document}